\documentclass[12pt,preprint]{aastex}

\begin{document} 

\title{Revealing the Jet Structure of GRB\,030329 with High Resolution
Multicolor Photometry}

\author{J.  Gorosabel\altaffilmark{1}, A.J.
Castro-Tirado\altaffilmark{1}, E. Ramirez-Ruiz\altaffilmark{2,3},
J. Granot\altaffilmark{4}, N. Caon\altaffilmark{5},
L.M. Cair\'os\altaffilmark{5}, E. Rubio-Herrera\altaffilmark{6},
S. Guziy\altaffilmark{1}, A. de Ugarte
Postigo\altaffilmark{1}, M. Jel\'{\i}nek\altaffilmark{1}}
\altaffiltext{1}{Instituto de Astrof\'{\i}sica de Andaluc\'{\i}a
(CSIC),  18008 Granada, Spain}
\altaffiltext{2}{Institute for Advanced Study, Einstein Drive,
Princeton NJ 08540, USA} \altaffiltext{3}{Chandra Fellow}
\altaffiltext{4}{KIPAC, Stanford University, Stanford CA 94309, USA} \altaffiltext{5}{Instituto de
Astrof\'{\i}sica de Canarias, 38205 La Laguna, Tenerife, Spain} 
\altaffiltext{6}{Astronomical Inst.  Anton Pannekoek, 1098 SJ Amsterdam, Netherlands}

\begin{abstract}

We present multicolor optical  observations of the nearby ($z=0.1685$)
GRB\,030329 obtained with the  same instrumentation over a time period
of  6 hours  for a  total of  an unprecedented  475 quasi-simultaneous
$B\,VR$ observations.  The achromatic steepening in the optical, which
occurs  at   $t  \sim0.7\;$days,  provides  evidence   for  a  dynamic
transition of the source, and  can be most readily explained by models
in which the  GRB ejecta are collimated into a  jet. Since the current
state-of-the-art   modeling  of   GRB  jets   is  still   flawed  with
uncertainties, we use these data  to critically assess some classes of
models that have been proposed in the literature. The data, especially
the smooth decline rate seen  in the optical afterglow, are consistent
with a model in which  GRB\,030329 was a homogeneous, sharp-edged jet,
viewed near  its edge interacting  with a uniform external  medium, or
viewed  near its symmetry  axis with  a stratified  wind-like external
environment. The  lack of short timescale fluctuations  in the optical
afterglow  flux  down  to  the  0.5  per  cent  level  puts  stringent
constraints on possible small scale angular inhomogeneities within the
jet or fluctuations in the external density.

\end{abstract}


\section{Introduction}

A  watershed  event occurred  on  29  March  2003 when  {\it  HETE-II}
localized rapidly  a long GRB \citep{Vand03}. The  prompt discovery of
the  fading optical  counterpart  \citep{tetal03,p03,u04,s03} combined
with its  exceptional brightness allowed  densely sampled observations
of this  afterglow (Lipkin  et al. 2004  and references  therein).  At
$z=0.1685$ \citep{g03a}, GRB\,030329 is  the third closest GRB to date
for  which an optical  afterglow (OA)  has been  discovered. It  is by
detailed study  of such  nearby events that  we have learned  the most
about the  range of physical processes relevant  to GRBs.  GRB\,030329
was the first burst with  a secure spectroscopic association to a type
Ic  supernova \citep[][]{stanek03,hjorth03,soko03} and  the monitoring
of its  afterglow polarization is  the best to date  \citep{g03b}.  In
its early  evolution, the  GRB\,030329 afterglow also  provided strong
evidence for slow shells with  modest Lorentz factors carrying most of
the kinetic energy in the relativistic ejecta \citep{gnp03}.

Follow-up observations within  $1.5\;$hr \citep{pp03} discovered an OA
with $R\sim12$, brighter than any  OA previously detected at a similar
epoch.   Soon  after, intensive  spectroscopic  monitoring revealed  a
supernova,  SN 2003dh,  with a  spectrum  very similar  to SN  1998bw.
These  observations showed  that $\sim0.4\;M_\sun$  of  $^{56}$Ni, the
parent nucleus of $^{56}$Co, was formed in the explosion, suggesting a
progenitor main  sequence mass  of 25-40 $M_\sun$  \citep{mazz03}. The
GRB\,030329 afterglow was monitored  in the radio, optical, and X-ray.
The radio  image was  resolved by the  VLBA \citep{t04},  its diameter
measured   to  be   $\sim0.2\;$pc  ($\sim0.5\;$pc)   after  $25\;$days
($83\;$days),  indicating an  average apparent  expansion  velocity of
$\sim5.7c$  ($\sim4.1c$).   This  decelerating  apparent  superluminal
expansion agrees  with expectations  of the standard  afterglow theory
\citep{o04}. The  image size  did not change  much between  $83\;$ and
$217\;$days \citep{t05}, favoring a  uniform external medium and a jet
with  little  lateral spreading.   The  GRB\,030329 lightcurve  indeed
showed a pan-chromatic steepening at  about half a day, which has been
attributed to a jet-like outflow \citep{p03}.

GRB\,030329  belongs to  a growing  group  of GRBs  for which  densely
monitored  OA lightcurves  show significant  deviations relative  to a
smooth     power-law    decay     -    most     notably    GRB\,021004
\citep{bersier,deugarte}   and   GRB\,011211   \citep{palli}.    These
deviations are interesting for  constraining the activity and identity
of the central  engine driving the GRB, but  complicate the simple jet
interpretation  mentioned  above.  In  this  {\it  Letter}, we  report
observations of the GRB\,030329 OA that overcome the previous sampling
limitations  with a  total of  an unprecedented  475  high resolution,
quasi-simultaneous  $B\,VR$-band observations  obtained with  the same
instrumentation over  a time period of  6 hours.  Here,  for the first
time, we establish a detailed multicolor optical lightcurve around the
break time and argue that in concert they provide good support for the
jet interpretation  of the lightcurve break.  The  paper is structured
as  follows.  \S~2 details  the observations  and data  analysis, \S~3
discusses   the  physical  implications,   and  our   conclusions  are
summarized in \S~4.

\section{Observations}

We commenced observations of  the GRB\,030329 OA, $9.44\;$hr after the
GRB, using the ALFOSC camera at the $2.5\;$m Nordic Optical Telescope.
We  continued monitoring  with the  same instrumentation  over  a time
period  of $6\;$hr.   The  data  were taken  in  sequential cycles  of
$B\,VR$-band images.  The monitoring is composed of 157 images in $B$,
158  in $V$,  and  160 in  the  $R$-band.  In  order  to enhance  time
resolution, the  2048$\times$2048 pixel chip  of ALFOSC was  binned by
2$\times$2 and  trimmed to a 800$\times$800 pixel  window. The typical
exposure  time per  frame  was $20\;$s,  achieving  a time  resolution
(including readout time) of about $\sim150\;$s between two consecutive
images taken  in the same band.   The photometry is  based on aperture
photometry running under IRAF.

\begin{figure}
\plotone{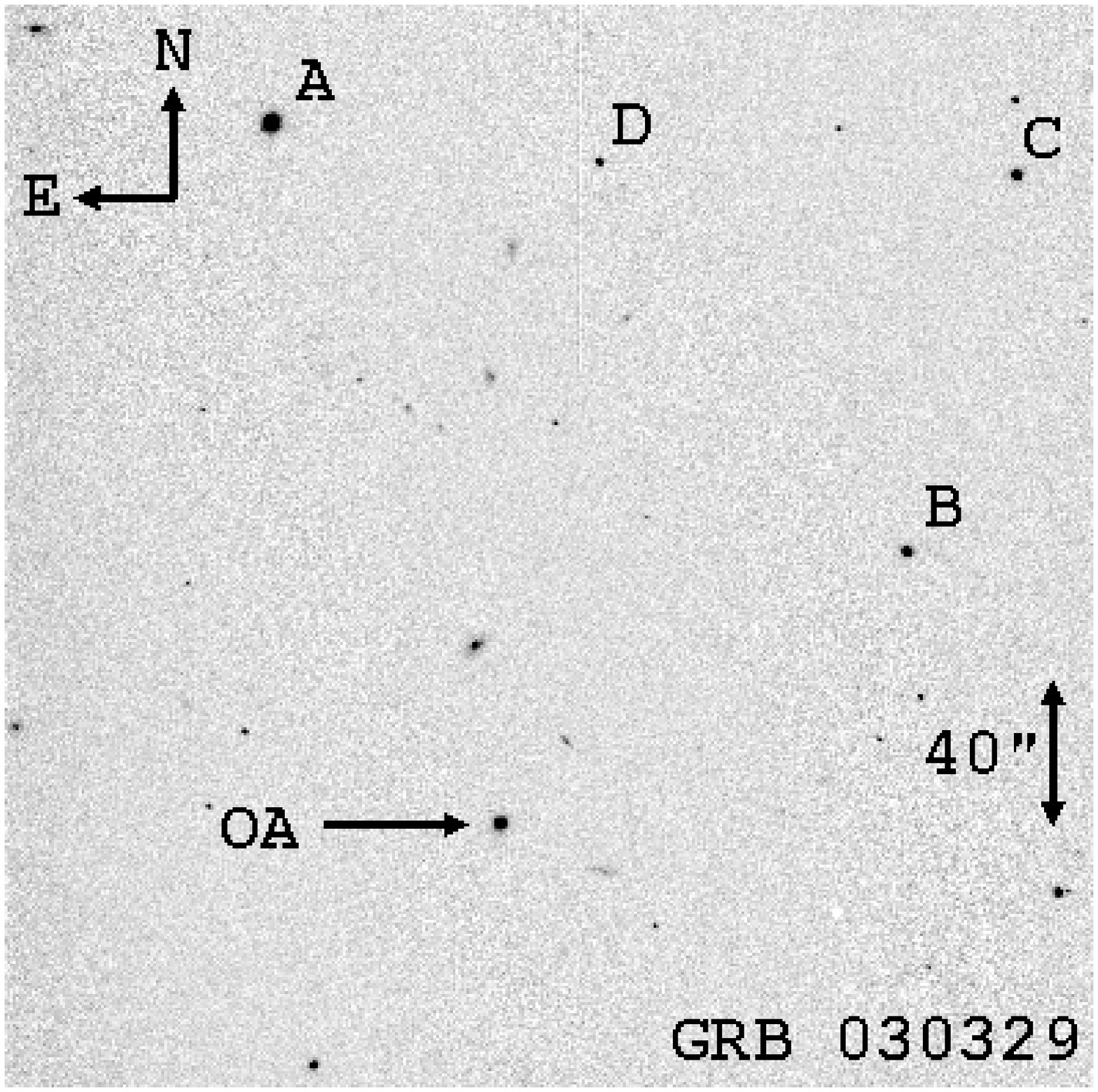}
\caption{{$B$-band  image  of  the   field  centered  on  the  optical
 afterglow of GRB\,030329. The reference  stars A,\,B,\,C,\,D and
 the OA are indicated.}}
\label{Fig1}
\end{figure}

\begin{figure}
\epsscale{0.8}
\plotone{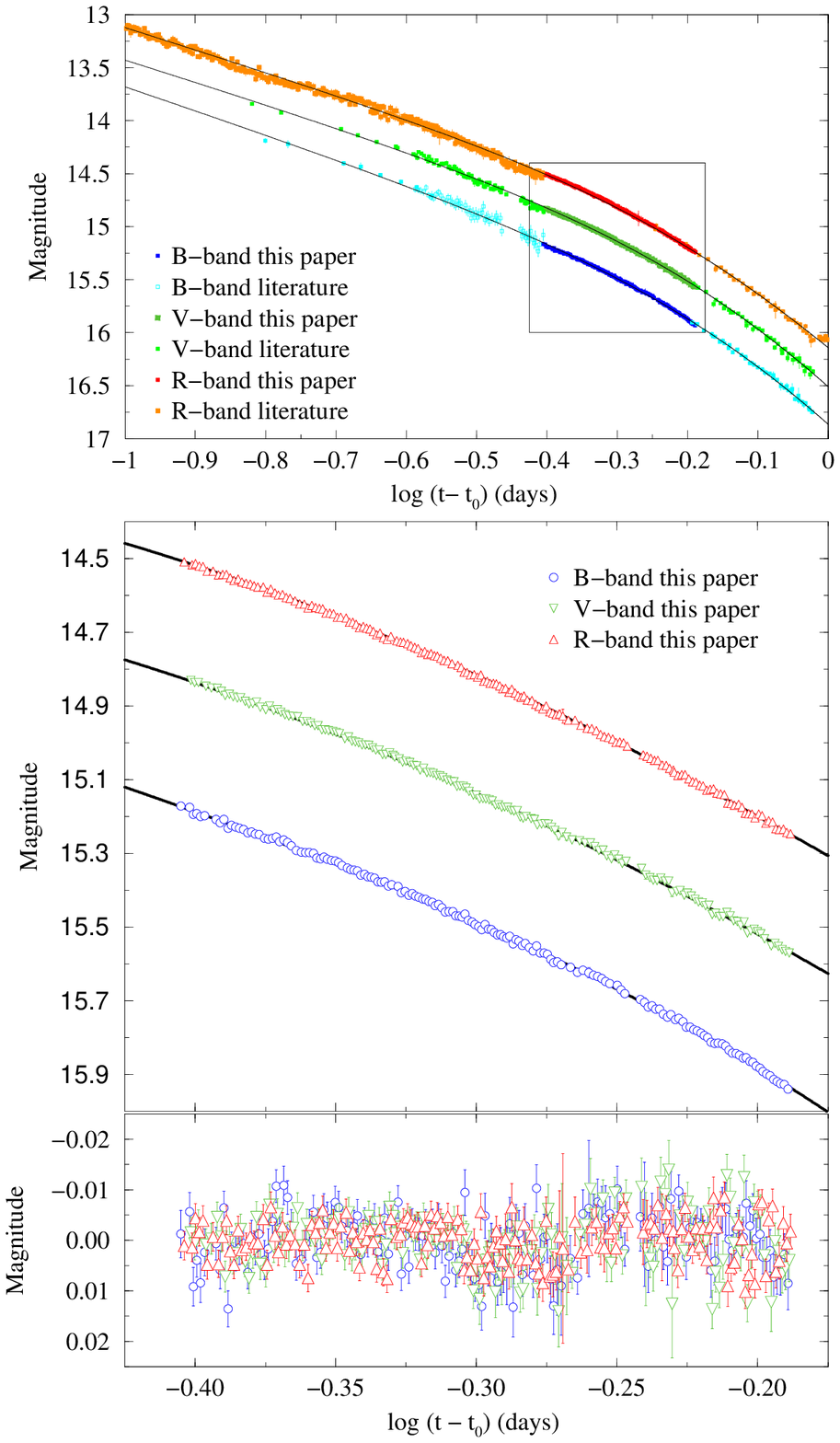}
\caption{{$B\,VR$-band lightcurves of the GRB\,030329 OA.  In addition 
to photometry from our group  ({\it middle panel}), we have  augmented
the  lightcurves with data  from  the literature  ({\it top  panel}).
The bottom panel shows  the $B\,VR$-band magnitude residuals with
respect to the corresponding Beuermann  fits (shown as 3 solid  curves
in the middle panel). }}
\label{Fig2}
\end{figure}

The  field of  view center  was selected  to include  the OA  and four
reference  stars  (A,  B,  C  and D,  see  Fig.~\ref{Fig1}).   The  OA
magnitude  ($R\sim15$) is  similar  to those  of  the reference  stars
\citep{Hend03},   showing   photometric   errors  (typically   $<1\%$)
comparable to  the OA.  Although  the airmass during  our observations
did  not reach  extreme values  ($Sec~{\rm z}  < 1.6$),  we introduced
airmass-dependent,   color   term  corrections   to   derive  the   OA
$B\,VR$-band  magnitudes.  This  correction is  more prominent  at the
$B$-band,  reaching  an  OA  flux  increment  up  to  $0.007\;$mag  at
$Sec~{\rm  z} =  1.6$.  For  calibration,  at a  given aperture,  four
different OA lightcurves were obtained per filter, each one based on a
different  reference star.   We  averaged the  residuals  of the  four
lightcurves,  instead  of averaging  over  the individual,  calibrated
lightcurves, so  that the error of  the residuals was  not affected by
the zero  point uncertainty of  the reference stars.  This  allowed to
decrease the  potential dispersion  contribution due to  the reference
stars'  statistical fluctuations.  This  method provided  for a  given
photometric aperture  the mean $B\,VR$-band fluctuations.  As a sanity
check, the process was repeated for 6 apertures, ranging from 1 to 2.5
times  the stars' FWHM.   The final  lightcurve and  the shape  of the
associated  residuals  (see  bottom  panel  of  Fig.~\ref{Fig2})  were
independent  of  the  adopted  aperture.  The  final  magnitudes  were
obtained averaging  the lightcurves yielded  for the 6  apertures (see
Table 1).

The $B$, $V$ and $R$-band lightcurves, along with numerous points from
other   groups   reported   in   the  literature,   are   plotted   in
Fig.~\ref{Fig2},  where it  is  evident that  the lightcurve  steepens
contemporaneously   in   all   bands,  between   $\sim0.3\;$days   and
$\sim0.9\;$days. In  order to characterize the shape  of the afterglow
lightcurve near  the jet break time,  $t_j$, we fit the  data with the
phenomenological \citet{beu99} function (Eq. 1);
$F(t,\nu )=F_{0}\ \nu ^{\beta }\ \left[
 (t/t_{j})^{-\alpha_{1}s}+(t/t_{j})^{-\alpha_{2}s}\right] ^{-1/s} + F_{host}(\nu)$,
 which provides  a good description of the  data, featuring a
smooth transition (whose sharpness is parametrized by $s$) between the
asymptotic power-law  indexes $\alpha_1$  and $\alpha_2$ at  early and
late   times,  respectively.    $F_{host}(\nu)$  corresponds   to  the
$BVR$-band  host galaxy  flux, which  was fixed  using  the magnitudes
reported by  \citet{Goro05}.  Fitting the $B$, $V$,  and $R$-band data
simultaneously  yields $t_j  = 0.72  \pm 0.2$  days, $s=1.6  \pm 0.7$,
$\beta=-0.89\pm  0.15$, $\alpha_1 =  -0.88 \pm  0.1$, and  $\alpha_2 =
-2.52  \pm   0.3$,  where  the  uncertainties   are  formal  1$\sigma$
errors.\footnote{Including   measurements   by   other   groups   (see
  Fig.~\ref{Fig2})  before  and after  our  observations  result in  a
  slightly sharper break, $s=2.3\pm0.3$.}  The $\chi^2$ for the fit is
acceptable: 446 for  638 degrees of freedom.  The  lightcurve shows no
significant  short timescale variability  on top  of the  smooth trend
described by the  functional fit above, down to  $\Delta B \sim0.005$,
$\Delta  V  \sim0.005$,  and   $\Delta  R  \sim0.005$  (i.e.   $\Delta
F/F\lesssim 0.5\%$) on timescales $\gtrsim 150\;$s ($\Delta t/t\gtrsim
10^{-2.5}$).  The  low $R$-band dispersion  agrees with that  found by
\citet{urat04}.  A linear fit to  $\beta$ vs.  $\log (t-t_0)$ yields a
slope  of  $-0.037  \pm  0.028$,  hence consistent  with  no  spectral
evolution between $9.44\;$hr and $15.56\;$hr after the burst.

\section{Discussion}

\subsection{Evidence for a jet in GRB\,030329}

A pan-chromatic steepening in the OA flux decay, like the one reported
here in  three optical bands ($B\,VR$) for  GRB\,030329, was predicted
for an outflow  collimated into a narrow jet  \citep{rhoads99}. Such a
``jet break'' occurs when the Lorentz factor, $\Gamma$, of the shocked
external medium  drops below $\theta_0^{-1}$, where  $\theta_0$ is the
initial half-opening  angle of the  jet \citep{sph99}.  At  this stage
the edge of  the jet becomes visible. In  principle, lateral expansion
of the jet might  also become important when $\Gamma\theta_0\sim1$, as
the edge of the jet and its center come into causal contact.  However,
hydrodynamic  simulations  \citep{gsim01}  show  a  modest  degree  of
lateral spreading as long as  the jet is relativistic, suggesting that
the jet break  in the lightcurve is caused  primarily because the edge
of the jet becomes visible.

At early times,  $t\ll t_j$, the lightcurve is  given by the spherical
solution,  where in  the  optical (depending  on  the density  profile
assumed for the  external matter) $\alpha_1 = 3(1-p)/4$  for a uniform
density  profile  and $\alpha_1=(1-3p)/4$  for  a  wind.  The  closure
relation  $\alpha_1-3\beta/2=0$  ($1/2$)  is  expected for  a  uniform
density  (wind).  Our  observed value,  $\alpha_1-3\beta/2  = -0.45\pm
0.3$, favors a uniform density, for which $p = 1+4\alpha_1/3 = 2.2 \pm
0.2$  $p =  1+2\beta =  2.78\pm 0.3$  are marginally  consistent.  For
relativistic  lateral spreading  of the  jet  in its  own rest  frame,
$\alpha_2=-p$  \citep{sph99}.   For little  or  no lateral  spreading,
$\alpha_1-\alpha_2=-2(d\log\Gamma/d\log  t)=(3-k)/(4-k)$  for a  power
law  external   density,  $\rho_{\rm  ext}\propto   r^{-k}$,  so  that
$\alpha_2=-3p/4$ for $k=0$. However, the value of $\alpha$ immediately
after the break is smaller than its asymptotic value\footnote{This has
  been shown  in numerical and  semi-analytic \citep{Rossi04} studies,
  and occurs because as the edge  of the jet becomes visible, we first
  start ``missing''  the outer part  of the afterglow image,  which is
  limb   brightened   \citep{gps99}.},   underestimating   $\alpha_2$.
Therefore,  our  measured  value  of  $\alpha_2  =  -2.52  \pm0.3$  is
consistent with this picture. For  $t_j = 0.72 \pm 0.2\;$days, we find
$\theta_0=0.097(E_{51}/n_0)^{-1/6}$
[$\theta_0=0.13(E_{51}/A_*)^{-1/2}$]  for  $k=0$  ($k=2$).  Using  the
constraints on  $E_{51}/n_0$ ($E_{51}/A_*$) from the  radio image size
\citep{grl05}, we derive\footnote{We take  into account an increase in
  energy by a factor of $\sim10$ due to refreshed shocks that occurred
  between   $t_j$  and  the   image  size   measurements.},  $\theta_0
\sim0.083-0.14$ $(\sim0.27 - 0.55)$.
 
The  densely monitored  observations presented  here provide  a unique
opportunity to  critically assess different jet models  that have been
proposed, and to constrain the  jet structure and the external density
profile.  We  start with a  uniform external medium, and  consider two
models  of  increasing  complexity.    First,  we  consider  a  simple
semi-analytical model \citep[model  1 of][]{gk03,rr05}, which neglects
lateral spreading.   This approximation is consistent  with results of
hydrodynamic  simulations  \citep{cgv04},  which are  also  considered
here.   The advantage  of the  hydrodynamic  model is  a reliable  and
rigorous treatment of the jet  dynamics, which provides insight on the
behavior  of  the  jet  and  lightcurves.  We  use  Eq.  1  to
determine the  sharpness parameter, $s$,  of the jet break.  While the
prompt  GRB seen  by all  observers within  the initial  jet aperture,
$\theta_{\rm obs} < \theta_0$, is the  same for a uniform jet, the jet
break is  sharper at smaller $\theta_{\rm obs}$  \citep{gsim01}. For a
uniform   jet,  we   find  $s=7.3,4.4,1.9,1.1,1.6$   for  $\theta_{\rm
  obs}=[0,{1 \over  4},{1 \over 2},{3 \over  4},1]\theta_0$. While the
basic  lightcurve  features  for  $\theta_{\rm obs}  <  \theta_0$  are
similar in  both models, the hydrodynamic model  gives sharper breaks;
e.g., we  find $s$ values ranging from\footnote{The  values given here
  are for  $p=2.2$, which is appropriate  for GRB\,030329.  Generally,
  we expect a  sharper jet break (larger $s$) for  large values of $p$
  \citep{Rossi04}.}  $s(\theta_{\rm  obs}=0) \sim10$ to $s(\theta_{\rm
  obs}=\theta_0)\sim1$.

For  a  wind  external  density  profile ($k=2$),  the  jet  break  is
smoother, because  both the Lorentz factor decreases  more slowly with
time \citep{kp00}  and the limb-brightening of the  afterglow image is
weaker  \citep{gl01}  when  compared  to  a  uniform  external  medium
($k=0$).  For  the uniform jet model,  we find $s=2.2,1.8,1.2,0.8,1.1$
for   $\theta_{\rm   obs}=[0,{1   \over   4},{1  \over   2},{3   \over
4},1]\theta_0$. While  the slow change in image  size observed between
$83\;$ and  $217\;$days favors a uniform  external medium \citep{t05},
the sharpness  change does  not.  The relatively  smooth break in
GRB\,030329,    with    $s=1.6\pm    0.7$,    suggests    $\theta_{\rm
obs}/\theta_0\sim0.7-1$ ($\lesssim 0.7$) for $k=0$ ($k=2$).  However,
a  large fraction  of OAs \citep{zeh}  show  evidence for sharper breaks, 
with $s$ values consistent with lower $\theta_{\rm obs}/\theta_0$ values 
for $k=0$, but are too sharp for any viewing angle for $k=2$.

\subsection{GRB\,030329 and its variability}

\begin{figure}
\epsscale{1.0}
\plotone{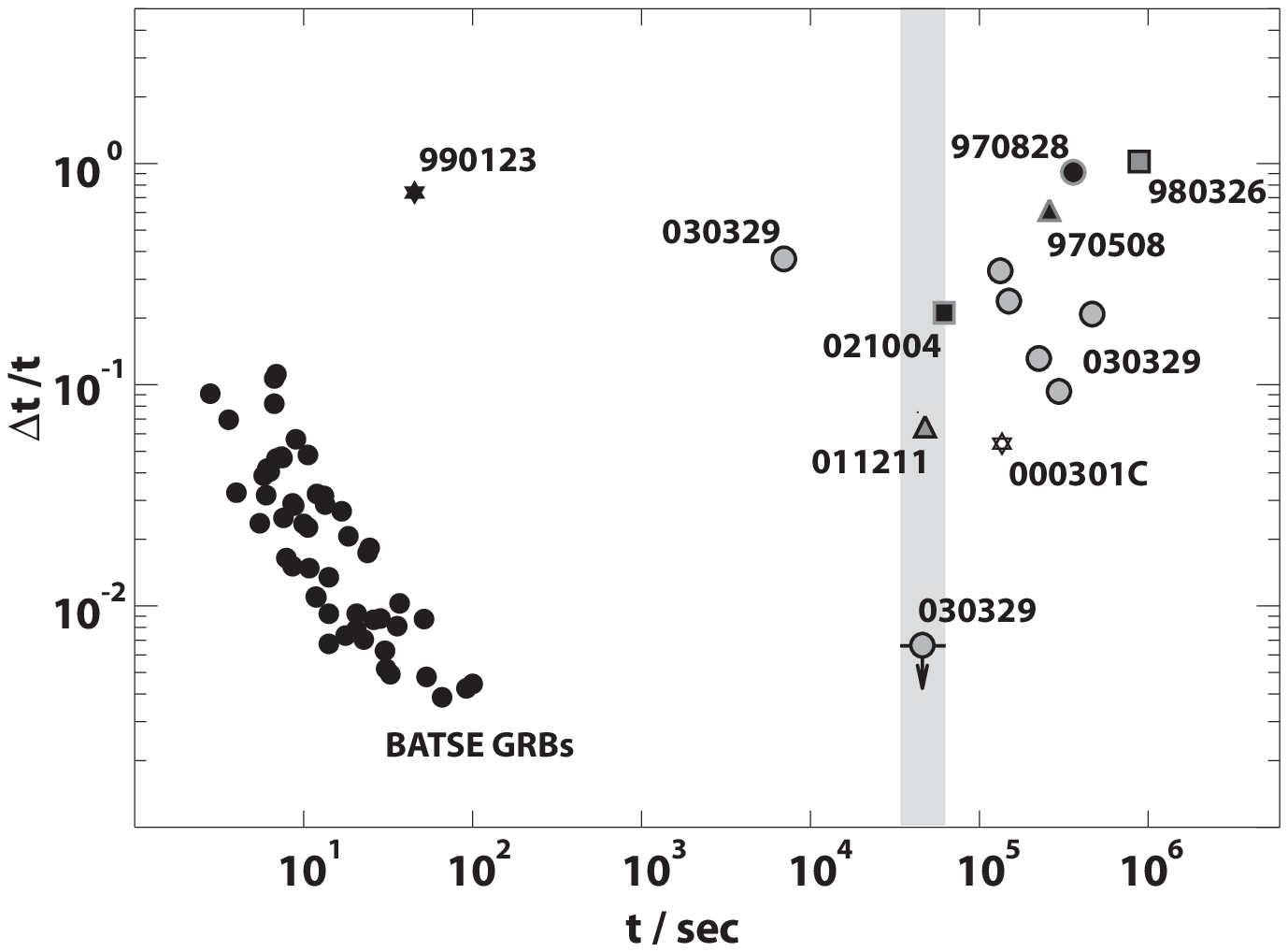}
\caption{{Evolution  of  the  variability,   $\Delta  t/  t$,  of  the
    GRB\,030329 OA (grey circle  at $t \lesssim 10^4\;$s, \citet{s03};
    circle  in  shaded  region,  this  paper; grey  circles  at  $t  >
    10^5\;$s,  \cite{lipkin04}).  Reported  variations  from a  simple
    power-law   decay   in  the   X-ray   afterglows  of   GRB\,970508
    \citep{piro98}  and  970828  \citep{yos98};  and  in  the  OAs  of
    GRB\,980326     \citep{bloom99},     GRB\,990123    \citep{rotse},
    GRB\,011211   \citep{palli},   GRB\,021004  \citep{bersier},   and
    GRB\,000301C \citep{gar} are  also illustrated.  The solid symbols
    are  long,  complex BATSE  GRBs  for  which  $\Delta t$  has  been
    estimated   from  the  width   of  the   autocorrelation  function
    \citep{fen99},  and   $t$  is  the  GRB   duration.   This  figure
    demonstrates  that while the  GRB\,030329 OA  displays substantial
    variability  at $t \gtrsim  10^5\;$s after  the burst,  we observe
    ({\it shaded  region}) no significantly variability,  $\Delta t/ t
    \lesssim  10^{-2.5}$, shortly  before the  onset of  the wave-like
    behavior. }}
\label{Fig3}
\end{figure}

GRB\,030329 belongs  to a  growing group of  bursts for  which densely
monitored  lightcurves  show  significant  ``bumps''  and  ``wiggles''
relative  to a  simple power-law  decay.  Fig.   \ref{Fig3}  shows the
temporal evolution of the variability timescale, $\Delta t/ t$, of the
various wiggles  observed in the GRB\,030329  lightcurve together with
other reported variations  from a simple power-law decay  in X-ray and
GRB OAs.  The short-term wave-like behavior of the GRB\,030329 optical
lightcurve  is  unprecedented.   The   early  OA  wiggle  seen  at  $t
\sim10^4\;$s  \citep{s03} could be  interpreted as  the result  of the
shock  wave  encountering  an  external  medium  of  variable  density
\citep[e.g.][]{wl00,rr01,laz02}.  The  pronounced bumps between  a day
and a week,  however, are most readily explained  by refreshed shocks,
i.e. slower shells  that were ejected from the source  near the end of
the GRB  and catch up with the  afterglow shock at a  later time, when
the latter  decelerates to slightly  below the shells'  Lorentz factor
\citep{gnp03}. These shells collide  with the shocked fluid behind the
afterglow shock and increase its energy \citep{rm98,p98,npg03,hp03}.

Fig.  \ref{Fig3} shows that while the OA shows substantial variability
at $t \gtrsim 10^5\;$s after the burst, we can set a very strict upper
limit on  the variability  shortly before the  onset of  the wave-like
behavior  of the  optical lightcurve,  of $\Delta  F/F\lesssim 5\times
10^{-3}$   on  timescales   $10^{-2.5}\lesssim  \Delta   t/t  \lesssim
0.25$. The low level of variability before one day constrains both the
possible angular inhomogeneity within  the afterglow shock and density
variations in the  external medium. Let us consider  variations in the
energy  per solid  angle $E_{\rm  iso}$  (external density  $n$) on  a
typical  angular  (length)  scale   $\theta_E$  ($l$)  which  cause  a
fractional  change $\Delta F_0/F_0$  in the  local emission  from each
such   small   region.\footnote{Since   in  our   case   $F_\nu\propto
  n^{1/2}E_{\rm  iso}^{(3+p)/4}$, a  larger fractional  change  in the
  density is needed  in order to produce the  same variation in flux.}
The total number of regions contributing to the total observed flux is
$N\sim(\Gamma\theta_E)^{-2}$   ($N\sim\Gamma^{-2}(R/l)^3$)  and  since
around   the    jet   break   time    $\Gamma\sim\theta_0^{-1}$   then
$N\sim(\theta_0/\theta_E)^2$ ($N\sim\theta_0^2(R/l)^3$). We expect the
fluctuations in the  total observed flux on a  timescale $\Delta t$ to
be  of  the order  $\Delta  F/F\sim N^{-1/2}(\Delta  t/t)^{1/2}(\Delta
F_0/F_0)$.   For our  limits this  implies  $N^{1/2}\gtrsim 100(\Delta
F_0/F_0)$,   i.e.    $l/R\lesssim   10^{-2}(\theta_0/0.1)^{2/3}(\Delta
F_0/F_0)^{-2/3}$         where         $R(t_j)\approx        5.4\times
10^{17}(E_{50}/n_0)\;$cm              or             $\theta_E\lesssim
10^{-3}(\theta_0/0.1)(\Delta F_0/F_0)^{-1}$.
 
\section{Conclusions}

We   report  on  high-resolution   photometric  observations   of  the
GRB\,030329  OA obtained  with the  same instrumentation  over  a time
period of 6  hours for a total of  475 quasi-simultaneous $B\,VR$-band
measurements  with   typical  statistical  photometric   errors  below
$0.005\;$mag.  Thus  our data, to  our knowledge, constitute  the most
complete  and  dense  sampling of  the  jet  like  behavior of  a  GRB
afterglow  to date.   Our well-sampled  multicolor lightcurve  is well
described by  a physical  model where the  ejecta are collimated  in a
jet.  We conclude that the observations, especially the smooth decline
rate seen in the OA, are  consistent with a model in which GRB\,030329
was a  sharp-edged GRB  jet.  We  consider both a  uniform and  a wind
density profile.  While both the  observed evolution of the image size
and the  relation between the  observed spectral and  temporal indexes
favor a  uniform medium,  both models are  consistent with  smooth jet
break in  our data.   Moreover, we derive  strict upper limits  on the
afterglow  short time  scale variability  during our  observations, of
$\lesssim 0.5\%$, which puts strong limits on possible fluctuations in
the  energy  per  solid angle  within  the  jet  and in  the  external
density. Evolving  photometric properties provide  a unique diagnostic
tool for GRB studies, and the complex lightcurve of the GRB\,030329 OA
emphasizes that data should be acquired with high sampling frequency.

\acknowledgments This work is  supported by IAS/NASA through a Chandra
Fellowship  award  PF3-40028  (ERR) under  contract  DE-AC03-76SF00515
(JG),  and  by  the  Spanish  Ministry  of  Science  through  programs
ESP2002-04124-C03-01 and AYA2004-01515.

\clearpage

\begin{deluxetable}{rrrrrrrr} 
\tablecolumns{3} 
\tablewidth{0pc} 
\tablecaption{To be published as an electronic table.}
\tablehead{ 
\colhead{MJD}    & \colhead{Band}   & \colhead{Magnitude}}
\startdata 
2728.3776157-2728.3777315&B&15.166$\pm$ 0.014\\
2728.3787153-2728.3788310&R&14.502$\pm$ 0.023\\
2728.3803472-2728.3805208&B&15.170$\pm$ 0.013\\
2728.3807870-2728.3809028&V&14.828$\pm$ 0.009\\
2728.3811690-2728.3812847&R&14.509$\pm$ 0.023\\
2728.3815394-2728.3817130&B&15.188$\pm$ 0.013\\
2728.3819792-2728.3820949&V&14.833$\pm$ 0.010\\
2728.3823495-2728.3824653&R&14.507$\pm$ 0.023\\
2728.3827199-2728.3828935&B&15.187$\pm$ 0.013\\
2728.3832292-2728.3833449&V&14.833$\pm$ 0.009\\
2728.3835995-2728.3837153&R&14.513$\pm$ 0.023\\
2728.3839583-2728.3841319&B&15.195$\pm$ 0.013\\
2728.3848032-2728.3849190&R&14.514$\pm$ 0.023\\
2728.3853125-2728.3854282&B&15.193$\pm$ 0.014\\
2728.3857639-2728.3858796&V&14.844$\pm$ 0.009\\
2728.3861227-2728.3862384&R&14.526$\pm$ 0.023\\
2728.3875463-2728.3876620&V&14.848$\pm$ 0.009\\
2728.3880671-2728.3881829&R&14.526$\pm$ 0.023\\
2728.3890162-2728.3891898&B&15.202$\pm$ 0.013\\
2728.3894676-2728.3895833&V&14.849$\pm$ 0.009\\
2728.3898727-2728.3899884&R&14.536$\pm$ 0.023\\
2728.3904630-2728.3906366&B&15.212$\pm$ 0.013\\
2728.3909144-2728.3910301&V&14.857$\pm$ 0.009\\
2728.3913079-2728.3914236&R&14.535$\pm$ 0.023\\
2728.3917361-2728.3919097&B&15.203$\pm$ 0.013\\
2728.3921875-2728.3923032&V&14.869$\pm$ 0.009\\
2728.3925926-2728.3927083&R&14.535$\pm$ 0.023\\
2728.3930208-2728.3931944&B&15.227$\pm$ 0.013\\
2728.3934606-2728.3935764&V&14.865$\pm$ 0.010\\
2728.3938657-2728.3939815&R&14.545$\pm$ 0.023\\
2728.3942593-2728.3944329&B&15.219$\pm$ 0.013\\
2728.3947222-2728.3948380&V&14.872$\pm$ 0.009\\
2728.3951273-2728.3952431&R&14.551$\pm$ 0.023\\
2728.3956134-2728.3957870&B&15.223$\pm$ 0.013\\
2728.3961343-2728.3962500&V&14.875$\pm$ 0.009\\
2728.3965394-2728.3966551&R&14.548$\pm$ 0.023\\
2728.3969329-2728.3971065&B&15.226$\pm$ 0.013\\
2728.3973843-2728.3975000&V&14.874$\pm$ 0.009\\
2728.3977778-2728.3978935&R&14.556$\pm$ 0.023\\
2728.3983333-2728.3985069&B&15.230$\pm$ 0.013\\
2728.3988310-2728.3989468&V&14.885$\pm$ 0.009\\
2728.3992245-2728.3993403&R&14.559$\pm$ 0.023\\
2728.3996644-2728.3998380&B&15.239$\pm$ 0.013\\
2728.4001736-2728.4002894&V&14.892$\pm$ 0.009\\
2728.4006134-2728.4007292&R&14.561$\pm$ 0.023\\
2728.4010301-2728.4012037&B&15.236$\pm$ 0.013\\
2728.4014931-2728.4016088&V&14.891$\pm$ 0.009\\
2728.4019097-2728.4020255&R&14.566$\pm$ 0.023\\
2728.4023032-2728.4024769&B&15.242$\pm$ 0.013\\
2728.4027662-2728.4028819&V&14.887$\pm$ 0.009\\
2728.4031829-2728.4032986&R&14.570$\pm$ 0.023\\
2728.4036227-2728.4037963&B&15.243$\pm$ 0.013\\
2728.4040856-2728.4042014&V&14.900$\pm$ 0.009\\
2728.4050000-2728.4051157&R&14.570$\pm$ 0.023\\
2728.4054167-2728.4055903&B&15.254$\pm$ 0.013\\
2728.4059144-2728.4060301&V&14.907$\pm$ 0.009\\
2728.4063310-2728.4064468&R&14.583$\pm$ 0.023\\
2728.4068171-2728.4069907&B&15.256$\pm$ 0.014\\
2728.4072801-2728.4073958&V&14.898$\pm$ 0.011\\
2728.4076736-2728.4077894&R&14.575$\pm$ 0.023\\
2728.4082176-2728.4083912&B&15.255$\pm$ 0.013\\
2728.4086690-2728.4087847&V&14.910$\pm$ 0.009\\
2728.4090741-2728.4091898&R&14.585$\pm$ 0.023\\
2728.4094676-2728.4096412&B&15.251$\pm$ 0.013\\
2728.4099306-2728.4100463&V&14.907$\pm$ 0.009\\
2728.4103241-2728.4104398&R&14.592$\pm$ 0.023\\
2728.4107292-2728.4109028&B&15.263$\pm$ 0.013\\
2728.4111921-2728.4113079&V&14.914$\pm$ 0.009\\
2728.4116667-2728.4117824&R&14.591$\pm$ 0.023\\
2728.4120949-2728.4122685&B&15.259$\pm$ 0.013\\
2728.4126273-2728.4127431&V&14.921$\pm$ 0.011\\
2728.4130208-2728.4131366&R&14.598$\pm$ 0.023\\
2728.4134375-2728.4136111&B&15.265$\pm$ 0.013\\
2728.4138773-2728.4139931&V&14.925$\pm$ 0.009\\
2728.4143634-2728.4144792&R&14.600$\pm$ 0.023\\
2728.4147685-2728.4149421&B&15.275$\pm$ 0.013\\
2728.4154167-2728.4155324&V&14.926$\pm$ 0.009\\
2728.4160764-2728.4161921&R&14.607$\pm$ 0.023\\
2728.4165741-2728.4167477&B&15.287$\pm$ 0.013\\
2728.4170718-2728.4171875&V&14.927$\pm$ 0.009\\
2728.4177083-2728.4178241&R&14.616$\pm$ 0.022\\
2728.4182755-2728.4184491&B&15.291$\pm$ 0.013\\
2728.4188310-2728.4189468&V&14.939$\pm$ 0.009\\
2728.4193750-2728.4194907&R&14.616$\pm$ 0.023\\
2728.4198264-2728.4200000&B&15.293$\pm$ 0.013\\
2728.4202778-2728.4203935&V&14.939$\pm$ 0.009\\
2728.4207523-2728.4208681&R&14.627$\pm$ 0.023\\
2728.4211458-2728.4213194&B&15.293$\pm$ 0.013\\
2728.4216782-2728.4217940&V&14.945$\pm$ 0.009\\
2728.4220833-2728.4221991&R&14.620$\pm$ 0.022\\
2728.4226389-2728.4228125&B&15.295$\pm$ 0.013\\
2728.4230903-2728.4232060&V&14.951$\pm$ 0.009\\
2728.4234722-2728.4235880&R&14.626$\pm$ 0.023\\
2728.4238657-2728.4240394&B&15.307$\pm$ 0.013\\
2728.4243403-2728.4244560&V&14.951$\pm$ 0.009\\
2728.4247801-2728.4248958&R&14.628$\pm$ 0.023\\
2728.4252083-2728.4253819&B&15.310$\pm$ 0.013\\
2728.4256713-2728.4257870&V&14.962$\pm$ 0.009\\
2728.4260764-2728.4261921&R&14.629$\pm$ 0.023\\
2728.4264815-2728.4266551&B&15.306$\pm$ 0.013\\
2728.4269213-2728.4270370&V&14.963$\pm$ 0.009\\
2728.4273148-2728.4274306&R&14.640$\pm$ 0.023\\
2728.4277199-2728.4278935&B&15.311$\pm$ 0.013\\
2728.4281944-2728.4283102&V&14.964$\pm$ 0.009\\
2728.4285764-2728.4286921&R&14.641$\pm$ 0.023\\
2728.4291667-2728.4293403&B&15.315$\pm$ 0.013\\
2728.4296181-2728.4297338&V&14.967$\pm$ 0.009\\
2728.4300347-2728.4301505&R&14.643$\pm$ 0.023\\
2728.4304977-2728.4306713&B&15.316$\pm$ 0.013\\
2728.4309375-2728.4310532&V&14.970$\pm$ 0.009\\
2728.4313310-2728.4314468&R&14.652$\pm$ 0.023\\
2728.4317245-2728.4318981&B&15.320$\pm$ 0.013\\
2728.4321644-2728.4322801&V&14.973$\pm$ 0.009\\
2728.4326042-2728.4327199&R&14.652$\pm$ 0.023\\
2728.4329977-2728.4331713&B&15.328$\pm$ 0.013\\
2728.4341898-2728.4343056&V&14.986$\pm$ 0.009\\
2728.4345949-2728.4347106&R&14.654$\pm$ 0.023\\
2728.4350116-2728.4351852&B&15.335$\pm$ 0.013\\
2728.4354630-2728.4355787&V&14.981$\pm$ 0.009\\
2728.4358449-2728.4359606&R&14.666$\pm$ 0.023\\
2728.4362731-2728.4364468&B&15.340$\pm$ 0.013\\
2728.4367130-2728.4368287&V&14.994$\pm$ 0.009\\
2728.4370949-2728.4372106&R&14.666$\pm$ 0.023\\
2728.4375347-2728.4377083&B&15.341$\pm$ 0.013\\
2728.4379745-2728.4380903&V&14.995$\pm$ 0.009\\
2728.4383681-2728.4384838&R&14.671$\pm$ 0.023\\
2728.4387616-2728.4389352&B&15.343$\pm$ 0.013\\
2728.4392130-2728.4393287&V&14.998$\pm$ 0.009\\
2728.4396065-2728.4397222&R&14.680$\pm$ 0.022\\
2728.4399884-2728.4401620&B&15.354$\pm$ 0.013\\
2728.4404745-2728.4405903&V&15.004$\pm$ 0.009\\
2728.4409491-2728.4410648&R&14.677$\pm$ 0.023\\
2728.4413657-2728.4415394&B&15.352$\pm$ 0.013\\
2728.4418171-2728.4419329&V&15.003$\pm$ 0.009\\
2728.4422106-2728.4423264&R&14.682$\pm$ 0.022\\
2728.4426042-2728.4427778&B&15.359$\pm$ 0.013\\
2728.4430787-2728.4431944&V&15.005$\pm$ 0.009\\
2728.4434838-2728.4435995&R&14.684$\pm$ 0.023\\
2728.4438773-2728.4440509&B&15.360$\pm$ 0.013\\
2728.4443519-2728.4444676&V&15.011$\pm$ 0.009\\
2728.4447917-2728.4449074&R&14.691$\pm$ 0.022\\
2728.4452199-2728.4453935&B&15.362$\pm$ 0.013\\
2728.4456597-2728.4457755&V&15.017$\pm$ 0.009\\
2728.4460532-2728.4461690&R&14.699$\pm$ 0.023\\
2728.4464699-2728.4466435&B&15.372$\pm$ 0.013\\
2728.4469213-2728.4470370&V&15.017$\pm$ 0.009\\
2728.4473032-2728.4474190&R&14.695$\pm$ 0.023\\
2728.4476968-2728.4478704&B&15.373$\pm$ 0.013\\
2728.4481713-2728.4482870&V&15.029$\pm$ 0.009\\
2728.4485764-2728.4486921&R&14.707$\pm$ 0.023\\
2728.4490278-2728.4492014&B&15.373$\pm$ 0.013\\
2728.4495139-2728.4496296&V&15.029$\pm$ 0.009\\
2728.4499769-2728.4500926&R&14.713$\pm$ 0.023\\
2728.4503704-2728.4505440&B&15.384$\pm$ 0.013\\
2728.4508796-2728.4509954&V&15.026$\pm$ 0.009\\
2728.4513542-2728.4514699&R&14.705$\pm$ 0.022\\
2728.4517708-2728.4519444&B&15.381$\pm$ 0.013\\
2728.4522222-2728.4523380&V&15.032$\pm$ 0.009\\
2728.4527894-2728.4529051&V&15.037$\pm$ 0.009\\
2728.4537269-2728.4538426&R&14.716$\pm$ 0.023\\
2728.4541204-2728.4542940&B&15.385$\pm$ 0.013\\
2728.4545949-2728.4547106&V&15.041$\pm$ 0.009\\
2728.4550347-2728.4551505&R&14.719$\pm$ 0.023\\
2728.4554167-2728.4555903&B&15.403$\pm$ 0.013\\
2728.4559028-2728.4560185&V&15.050$\pm$ 0.009\\
2728.4563194-2728.4564352&R&14.720$\pm$ 0.022\\
2728.4568056-2728.4569792&B&15.399$\pm$ 0.013\\
2728.4572685-2728.4573843&V&15.048$\pm$ 0.009\\
2728.4578356-2728.4579514&R&14.724$\pm$ 0.022\\
2728.4582639-2728.4584375&B&15.410$\pm$ 0.013\\
2728.4587153-2728.4588310&V&15.052$\pm$ 0.009\\
2728.4591435-2728.4592593&R&14.729$\pm$ 0.022\\
2728.4595602-2728.4597338&B&15.405$\pm$ 0.013\\
2728.4601620-2728.4602778&V&15.059$\pm$ 0.009\\
2728.4605556-2728.4606713&R&14.734$\pm$ 0.022\\
2728.4609954-2728.4611690&B&15.411$\pm$ 0.013\\
2728.4614468-2728.4615625&V&15.061$\pm$ 0.009\\
2728.4618866-2728.4620023&R&14.738$\pm$ 0.023\\
2728.4623148-2728.4624884&B&15.415$\pm$ 0.013\\
2728.4627662-2728.4628819&V&15.071$\pm$ 0.009\\
2728.4632292-2728.4633449&R&14.749$\pm$ 0.023\\
2728.4636574-2728.4638310&B&15.419$\pm$ 0.013\\
2728.4641088-2728.4642245&V&15.067$\pm$ 0.009\\
2728.4644907-2728.4646065&R&14.745$\pm$ 0.023\\
2728.4649306-2728.4651042&B&15.421$\pm$ 0.013\\
2728.4653935-2728.4655093&V&15.078$\pm$ 0.009\\
2728.4657755-2728.4658912&R&14.753$\pm$ 0.023\\
2728.4661806-2728.4663542&B&15.427$\pm$ 0.013\\
2728.4666551-2728.4667708&V&15.079$\pm$ 0.009\\
2728.4670255-2728.4671412&R&14.751$\pm$ 0.023\\
2728.4674421-2728.4676157&B&15.428$\pm$ 0.013\\
2728.4678704-2728.4679861&V&15.081$\pm$ 0.009\\
2728.4682870-2728.4684028&R&14.757$\pm$ 0.022\\
2728.4686921-2728.4688657&B&15.437$\pm$ 0.013\\
2728.4691667-2728.4692824&V&15.087$\pm$ 0.009\\
2728.4695949-2728.4697106&R&14.759$\pm$ 0.023\\
2728.4700463-2728.4702199&B&15.435$\pm$ 0.013\\
2728.4704861-2728.4706019&V&15.088$\pm$ 0.009\\
2728.4708796-2728.4709954&R&14.764$\pm$ 0.023\\
2728.4712847-2728.4714583&B&15.443$\pm$ 0.013\\
2728.4717477-2728.4718634&V&15.092$\pm$ 0.010\\
2728.4721412-2728.4722569&R&14.770$\pm$ 0.023\\
2728.4726273-2728.4728009&B&15.454$\pm$ 0.013\\
2728.4733449-2728.4734606&V&15.099$\pm$ 0.009\\
2728.4737269-2728.4738426&R&14.771$\pm$ 0.022\\
2728.4742477-2728.4744213&B&15.454$\pm$ 0.013\\
2728.4749074-2728.4750231&V&15.103$\pm$ 0.009\\
2728.4753125-2728.4754282&R&14.779$\pm$ 0.023\\
2728.4757060-2728.4758796&B&15.454$\pm$ 0.013\\
2728.4761458-2728.4762616&V&15.104$\pm$ 0.009\\
2728.4769444-2728.4771181&B&15.465$\pm$ 0.013\\
2728.4774074-2728.4775231&V&15.117$\pm$ 0.009\\
2728.4778125-2728.4779282&R&14.784$\pm$ 0.022\\
2728.4781944-2728.4783681&B&15.466$\pm$ 0.013\\
2728.4786343-2728.4787500&V&15.115$\pm$ 0.009\\
2728.4790509-2728.4791667&R&14.790$\pm$ 0.023\\
2728.4794444-2728.4796181&B&15.468$\pm$ 0.013\\
2728.4798958-2728.4800116&V&15.116$\pm$ 0.009\\
2728.4802778-2728.4803935&R&14.801$\pm$ 0.023\\
2728.4806713-2728.4808449&B&15.460$\pm$ 0.014\\
2728.4811227-2728.4812384&V&15.121$\pm$ 0.009\\
2728.4815162-2728.4816319&R&14.802$\pm$ 0.023\\
2728.4819097-2728.4820833&B&15.477$\pm$ 0.013\\
2728.4823958-2728.4825116&V&15.130$\pm$ 0.010\\
2728.4828819-2728.4829977&R&14.809$\pm$ 0.023\\
2728.4834259-2728.4835995&B&15.482$\pm$ 0.014\\
2728.4839815-2728.4840972&V&15.141$\pm$ 0.010\\
2728.4843866-2728.4845023&R&14.808$\pm$ 0.023\\
2728.4847801-2728.4849537&B&15.489$\pm$ 0.014\\
2728.4852546-2728.4853704&V&15.140$\pm$ 0.011\\
2728.4856829-2728.4857986&R&14.815$\pm$ 0.024\\
2728.4862500-2728.4864236&B&15.495$\pm$ 0.014\\
2728.4867014-2728.4868171&V&15.150$\pm$ 0.012\\
2728.4871296-2728.4872454&R&14.810$\pm$ 0.024\\
2728.4875347-2728.4877083&B&15.503$\pm$ 0.014\\
2728.4880093-2728.4881250&V&15.149$\pm$ 0.011\\
2728.4883912-2728.4885069&R&14.827$\pm$ 0.023\\
2728.4887847-2728.4889583&B&15.488$\pm$ 0.014\\
2728.4892361-2728.4893519&V&15.150$\pm$ 0.010\\
2728.4896644-2728.4897801&R&14.827$\pm$ 0.023\\
2728.4901389-2728.4903125&B&15.502$\pm$ 0.014\\
2728.4905903-2728.4907060&V&15.153$\pm$ 0.011\\
2728.4909954-2728.4911111&R&14.833$\pm$ 0.024\\
2728.4914120-2728.4915856&B&15.502$\pm$ 0.014\\
2728.4918519-2728.4919676&V&15.156$\pm$ 0.010\\
2728.4922454-2728.4923611&R&14.832$\pm$ 0.024\\
2728.4926505-2728.4928241&B&15.511$\pm$ 0.013\\
2728.4931366-2728.4932523&V&15.168$\pm$ 0.012\\
2728.4935764-2728.4936921&R&14.833$\pm$ 0.023\\
2728.4940162-2728.4941898&B&15.517$\pm$ 0.014\\
2728.4944792-2728.4945949&V&15.166$\pm$ 0.010\\
2728.4948727-2728.4949884&R&14.840$\pm$ 0.023\\
2728.4952894-2728.4954630&B&15.516$\pm$ 0.016\\
2728.4957523-2728.4958681&V&15.173$\pm$ 0.013\\
2728.4961343-2728.4962500&R&14.847$\pm$ 0.025\\
2728.4965509-2728.4967245&B&15.522$\pm$ 0.014\\
2728.4969907-2728.4971065&V&15.172$\pm$ 0.012\\
2728.4973958-2728.4975116&R&14.846$\pm$ 0.027\\
2728.4978241-2728.4979977&B&15.525$\pm$ 0.018\\
2728.4983218-2728.4984375&V&15.173$\pm$ 0.013\\
2728.4987963-2728.4989120&R&14.857$\pm$ 0.024\\
2728.4991898-2728.4993634&B&15.520$\pm$ 0.014\\
2728.4996759-2728.4997917&V&15.181$\pm$ 0.013\\
2728.5000810-2728.5001968&R&14.848$\pm$ 0.026\\
2728.5005787-2728.5007523&B&15.541$\pm$ 0.015\\
2728.5010185-2728.5011343&V&15.184$\pm$ 0.014\\
2728.5014236-2728.5015394&R&14.859$\pm$ 0.027\\
2728.5018056-2728.5019792&B&15.530$\pm$ 0.015\\
2728.5022685-2728.5023843&V&15.182$\pm$ 0.010\\
2728.5026736-2728.5027894&R&14.866$\pm$ 0.023\\
2728.5030903-2728.5032639&B&15.534$\pm$ 0.014\\
2728.5035301-2728.5036458&V&15.199$\pm$ 0.010\\
2728.5039815-2728.5040972&R&14.870$\pm$ 0.023\\
2728.5044329-2728.5046065&B&15.546$\pm$ 0.013\\
2728.5048958-2728.5050116&V&15.202$\pm$ 0.010\\
2728.5052778-2728.5053935&R&14.867$\pm$ 0.023\\
2728.5056713-2728.5058449&B&15.540$\pm$ 0.014\\
2728.5061111-2728.5062269&V&15.198$\pm$ 0.010\\
2728.5067014-2728.5068171&R&14.878$\pm$ 0.023\\
2728.5075231-2728.5076968&B&15.552$\pm$ 0.013\\
2728.5081944-2728.5083102&V&15.205$\pm$ 0.010\\
2728.5086921-2728.5088079&R&14.879$\pm$ 0.023\\
2728.5091319-2728.5093056&B&15.560$\pm$ 0.014\\
2728.5096991-2728.5098148&V&15.206$\pm$ 0.010\\
2728.5101389-2728.5102546&R&14.889$\pm$ 0.025\\
2728.5105440-2728.5107176&B&15.547$\pm$ 0.014\\
2728.5110069-2728.5111227&V&15.221$\pm$ 0.013\\
2728.5116898-2728.5118056&R&14.896$\pm$ 0.026\\
2728.5121296-2728.5123032&B&15.565$\pm$ 0.014\\
2728.5127894-2728.5129051&V&15.219$\pm$ 0.012\\
2728.5132407-2728.5133565&R&14.902$\pm$ 0.024\\
2728.5138079-2728.5139815&B&15.566$\pm$ 0.014\\
2728.5142824-2728.5143981&V&15.218$\pm$ 0.011\\
2728.5147106-2728.5148264&R&14.894$\pm$ 0.026\\
2728.5152315-2728.5154051&B&15.570$\pm$ 0.015\\
2728.5156944-2728.5158102&V&15.219$\pm$ 0.012\\
2728.5163542-2728.5164699&R&14.910$\pm$ 0.023\\
2728.5167708-2728.5169444&B&15.583$\pm$ 0.014\\
2728.5172106-2728.5173264&V&15.228$\pm$ 0.011\\
2728.5176273-2728.5177431&R&14.914$\pm$ 0.025\\
2728.5180903-2728.5182639&B&15.592$\pm$ 0.015\\
2728.5185995-2728.5187153&V&15.236$\pm$ 0.013\\
2728.5191088-2728.5192245&R&14.918$\pm$ 0.024\\
2728.5197106-2728.5198843&B&15.590$\pm$ 0.021\\
2728.5201736-2728.5202894&V&15.252$\pm$ 0.016\\
2728.5206829-2728.5207986&R&14.915$\pm$ 0.035\\
2728.5212269-2728.5214005&B&15.597$\pm$ 0.021\\
2728.5216667-2728.5217824&V&15.242$\pm$ 0.023\\
2728.5220718-2728.5221875&R&14.922$\pm$ 0.068\\
2728.5235995-2728.5238310&V&15.254$\pm$ 0.021\\
2728.5242824-2728.5246296&R&14.933$\pm$ 0.026\\
2728.5251389-2728.5256597&B&15.604$\pm$ 0.014\\
2728.5260069-2728.5263542&V&15.262$\pm$ 0.009\\
2728.5274190-2728.5277662&R&14.933$\pm$ 0.027\\
2728.5284491-2728.5289699&B&15.618$\pm$ 0.013\\
2728.5293056-2728.5296528&V&15.264$\pm$ 0.009\\
2728.5300231-2728.5303704&R&14.947$\pm$ 0.023\\
2728.5307639-2728.5312847&B&15.612$\pm$ 0.014\\
2728.5317245-2728.5320718&V&15.270$\pm$ 0.010\\
2728.5327546-2728.5329861&R&14.954$\pm$ 0.025\\
2728.5335995-2728.5339468&B&15.619$\pm$ 0.032\\
2728.5342477-2728.5344792&V&15.270$\pm$ 0.015\\
2728.5347917-2728.5350231&R&14.956$\pm$ 0.024\\
2728.5355093-2728.5358565&B&15.626$\pm$ 0.013\\
2728.5362616-2728.5364931&V&15.281$\pm$ 0.010\\
2728.5369444-2728.5371759&R&14.961$\pm$ 0.023\\
2728.5375347-2728.5378819&B&15.630$\pm$ 0.014\\
2728.5381481-2728.5383796&V&15.290$\pm$ 0.011\\
2728.5388079-2728.5390394&R&14.972$\pm$ 0.023\\
2728.5393866-2728.5397338&B&15.635$\pm$ 0.013\\
2728.5400926-2728.5403241&V&15.289$\pm$ 0.010\\
2728.5406366-2728.5408681&R&14.977$\pm$ 0.026\\
2728.5411343-2728.5414815&B&15.644$\pm$ 0.016\\
2728.5418056-2728.5420370&V&15.306$\pm$ 0.012\\
2728.5424421-2728.5426736&R&14.979$\pm$ 0.024\\
2728.5429977-2728.5433449&B&15.648$\pm$ 0.013\\
2728.5439120-2728.5441435&V&15.313$\pm$ 0.010\\
2728.5444213-2728.5446528&R&14.984$\pm$ 0.023\\
2728.5453704-2728.5456019&V&15.303$\pm$ 0.009\\
2728.5459144-2728.5461458&R&14.993$\pm$ 0.023\\
2728.5464352-2728.5467824&B&15.654$\pm$ 0.014\\
2728.5470718-2728.5473032&V&15.314$\pm$ 0.010\\
2728.5476852-2728.5479167&R&14.992$\pm$ 0.023\\
2728.5482176-2728.5485648&B&15.665$\pm$ 0.013\\
2728.5488426-2728.5490741&V&15.326$\pm$ 0.010\\
2728.5493750-2728.5496065&R&14.995$\pm$ 0.023\\
2728.5499074-2728.5502546&B&15.676$\pm$ 0.013\\
2728.5506366-2728.5508681&V&15.320$\pm$ 0.010\\
2728.5512269-2728.5514583&R&15.003$\pm$ 0.025\\
2728.5571644-2728.5575116&B&15.691$\pm$ 0.017\\
2728.5578819-2728.5581134&V&15.339$\pm$ 0.014\\
2728.5586458-2728.5588773&R&15.027$\pm$ 0.029\\
2728.5591782-2728.5595255&B&15.699$\pm$ 0.026\\
2728.5598032-2728.5600347&V&15.357$\pm$ 0.027\\
2728.5603241-2728.5605556&R&15.031$\pm$ 0.025\\
2728.5608565-2728.5612037&B&15.713$\pm$ 0.014\\
2728.5617245-2728.5619560&V&15.365$\pm$ 0.010\\
2728.5622222-2728.5624537&R&15.036$\pm$ 0.023\\
2728.5627315-2728.5630787&B&15.709$\pm$ 0.014\\
2728.5633681-2728.5635995&V&15.371$\pm$ 0.012\\
2728.5639005-2728.5641319&R&15.042$\pm$ 0.024\\
2728.5644213-2728.5647685&B&15.718$\pm$ 0.014\\
2728.5650463-2728.5652778&V&15.372$\pm$ 0.010\\
2728.5655671-2728.5657986&R&15.042$\pm$ 0.024\\
2728.5663079-2728.5666551&B&15.717$\pm$ 0.014\\
2728.5669560-2728.5671875&V&15.365$\pm$ 0.012\\
2728.5675810-2728.5678125&R&15.055$\pm$ 0.036\\
2728.5681597-2728.5685069&B&15.731$\pm$ 0.014\\
2728.5688657-2728.5690972&V&15.373$\pm$ 0.013\\
2728.5694444-2728.5696759&R&15.055$\pm$ 0.024\\
2728.5700810-2728.5704282&B&15.739$\pm$ 0.014\\
2728.5707176-2728.5709491&V&15.373$\pm$ 0.014\\
2728.5712269-2728.5714583&R&15.063$\pm$ 0.024\\
2728.5717130-2728.5720602&B&15.732$\pm$ 0.015\\
2728.5723264-2728.5725579&V&15.405$\pm$ 0.024\\
2728.5729051-2728.5730208&R&15.069$\pm$ 0.026\\
2728.5732986-2728.5735301&R&15.073$\pm$ 0.024\\
2728.5738773-2728.5742245&B&15.746$\pm$ 0.013\\
2728.5745139-2728.5747454&V&15.401$\pm$ 0.011\\
2728.5750694-2728.5753009&R&15.083$\pm$ 0.024\\
2728.5756134-2728.5759606&B&15.743$\pm$ 0.013\\
2728.5762731-2728.5765046&V&15.399$\pm$ 0.009\\
2728.5769560-2728.5771875&R&15.080$\pm$ 0.023\\
2728.5774884-2728.5778356&B&15.752$\pm$ 0.013\\
2728.5781597-2728.5783912&V&15.411$\pm$ 0.009\\
2728.5786690-2728.5789005&R&15.081$\pm$ 0.023\\
2728.5792014-2728.5795486&B&15.767$\pm$ 0.014\\
2728.5798611-2728.5800926&V&15.406$\pm$ 0.011\\
2728.5803819-2728.5806134&R&15.092$\pm$ 0.023\\
2728.5809491-2728.5812963&B&15.766$\pm$ 0.013\\
2728.5817708-2728.5820023&V&15.414$\pm$ 0.009\\
2728.5823148-2728.5825463&R&15.100$\pm$ 0.023\\
2728.5828704-2728.5832176&B&15.773$\pm$ 0.013\\
2728.5834838-2728.5837153&V&15.423$\pm$ 0.009\\
2728.5839699-2728.5842014&R&15.101$\pm$ 0.022\\
2728.5845833-2728.5849306&B&15.776$\pm$ 0.013\\
2728.5852199-2728.5854514&V&15.430$\pm$ 0.009\\
2728.5857292-2728.5859606&R&15.109$\pm$ 0.022\\
2728.5862616-2728.5866088&B&15.782$\pm$ 0.013\\
2728.5869329-2728.5871644&V&15.432$\pm$ 0.009\\
2728.5874421-2728.5876736&R&15.111$\pm$ 0.022\\
2728.5880208-2728.5883681&B&15.790$\pm$ 0.013\\
2728.5886343-2728.5888657&V&15.440$\pm$ 0.009\\
2728.5892245-2728.5894560&R&15.122$\pm$ 0.023\\
2728.5897685-2728.5901157&B&15.794$\pm$ 0.014\\
2728.5903935-2728.5906250&V&15.439$\pm$ 0.012\\
2728.5908912-2728.5911227&R&15.120$\pm$ 0.023\\
2728.5913773-2728.5917245&B&15.807$\pm$ 0.014\\
2728.5920718-2728.5923032&V&15.463$\pm$ 0.010\\
2728.5925694-2728.5928009&R&15.120$\pm$ 0.024\\
2728.5931134-2728.5934606&B&15.810$\pm$ 0.014\\
2728.5937847-2728.5940162&V&15.461$\pm$ 0.010\\
2728.5943519-2728.5945833&R&15.130$\pm$ 0.024\\
2728.5949537-2728.5953009&B&15.810$\pm$ 0.014\\
2728.5956481-2728.5958796&V&15.469$\pm$ 0.010\\
2728.5962384-2728.5964699&R&15.132$\pm$ 0.024\\
2728.5969213-2728.5972685&B&15.812$\pm$ 0.014\\
2728.5975231-2728.5977546&V&15.469$\pm$ 0.010\\
2728.5980671-2728.5982986&R&15.135$\pm$ 0.023\\
2728.5986111-2728.5989583&B&15.817$\pm$ 0.014\\
2728.5993056-2728.5995370&V&15.457$\pm$ 0.011\\
2728.5998032-2728.6000347&R&15.158$\pm$ 0.024\\
2728.6003819-2728.6007292&B&15.828$\pm$ 0.014\\
2728.6010417-2728.6012731&V&15.467$\pm$ 0.010\\
2728.6015278-2728.6017593&R&15.159$\pm$ 0.023\\
2728.6020718-2728.6024190&B&15.836$\pm$ 0.013\\
2728.6026852-2728.6029167&V&15.481$\pm$ 0.009\\
2728.6032176-2728.6034491&R&15.163$\pm$ 0.022\\
2728.6039931-2728.6043403&B&15.838$\pm$ 0.013\\
2728.6046644-2728.6048958&V&15.482$\pm$ 0.009\\
2728.6051736-2728.6054051&R&15.161$\pm$ 0.022\\
2728.6056713-2728.6060185&B&15.841$\pm$ 0.013\\
2728.6063310-2728.6065625&V&15.493$\pm$ 0.010\\
2728.6069560-2728.6071875&R&15.169$\pm$ 0.023\\
2728.6074884-2728.6078356&B&15.850$\pm$ 0.013\\
2728.6081134-2728.6083449&V&15.483$\pm$ 0.011\\
2728.6086343-2728.6088657&R&15.184$\pm$ 0.024\\
2728.6091435-2728.6094907&B&15.849$\pm$ 0.014\\
2728.6098148-2728.6100463&V&15.513$\pm$ 0.010\\
2728.6103935-2728.6106250&R&15.186$\pm$ 0.023\\
2728.6109144-2728.6112616&B&15.858$\pm$ 0.013\\
2728.6115509-2728.6117824&V&15.508$\pm$ 0.010\\
2728.6120486-2728.6122801&R&15.187$\pm$ 0.023\\
2728.6125810-2728.6129282&B&15.861$\pm$ 0.013\\
2728.6132292-2728.6134606&V&15.503$\pm$ 0.010\\
2728.6137384-2728.6139699&R&15.197$\pm$ 0.023\\
2728.6142940-2728.6146412&B&15.872$\pm$ 0.013\\
2728.6150116-2728.6152431&V&15.509$\pm$ 0.010\\
2728.6155556-2728.6157870&R&15.187$\pm$ 0.023\\
2728.6160995-2728.6164468&B&15.876$\pm$ 0.013\\
2728.6167477-2728.6169792&V&15.515$\pm$ 0.009\\
2728.6172569-2728.6174884&R&15.191$\pm$ 0.023\\
2728.6177778-2728.6181250&B&15.886$\pm$ 0.013\\
2728.6185648-2728.6187963&V&15.529$\pm$ 0.010\\
2728.6191667-2728.6193981&R&15.211$\pm$ 0.023\\
2728.6196875-2728.6200347&B&15.892$\pm$ 0.013\\
2728.6208796-2728.6211111&R&15.205$\pm$ 0.023\\
2728.6214931-2728.6218403&B&15.897$\pm$ 0.013\\
2728.6221644-2728.6223958&V&15.548$\pm$ 0.014\\
2728.6228009-2728.6230324&R&15.208$\pm$ 0.023\\
2728.6233449-2728.6236921&B&15.908$\pm$ 0.014\\
2728.6239931-2728.6242245&V&15.541$\pm$ 0.010\\
2728.6245370-2728.6247685&R&15.226$\pm$ 0.023\\
2728.6251505-2728.6254977&B&15.912$\pm$ 0.014\\
2728.6257986-2728.6260301&V&15.549$\pm$ 0.011\\
2728.6263194-2728.6265509&R&15.228$\pm$ 0.024\\
2728.6269792-2728.6273264&B&15.919$\pm$ 0.013\\
2728.6277431-2728.6279745&V&15.559$\pm$ 0.010\\
2728.6283681-2728.6285995&R&15.227$\pm$ 0.023\\
2728.6289583-2728.6293056&B&15.921$\pm$ 0.013\\
2728.6296065-2728.6298380&V&15.563$\pm$ 0.011\\
2728.6302778-2728.6305093&R&15.236$\pm$ 0.024\\
2728.6308681-2728.6312153&B&15.933$\pm$ 0.014\\
2728.6315162-2728.6317477&V&15.567$\pm$ 0.010\\
2728.6322569-2728.6324884&R&15.239$\pm$ 0.024\\
\enddata 
\end{deluxetable}
\end{document}